\documentclass[prl,floatfix,superscriptaddress,amsmath,amssymb,twocolumn]{revtex4}
\usepackage[latin1]{inputenc}
\usepackage{amsmath}
\usepackage{amssymb}
\usepackage{graphicx}
\usepackage{graphics}
\usepackage{color}
\usepackage{multirow}

\def\draft{\oddsidemargin -.5truein
	\def\@oddfoot{\sl preliminary draft \hfil
	\rm\thepage\hfil\sl\today\quad\militarytime}
	\let\@evenfoot\@oddfoot \overfullrule 3pt
	\let\label=\draftlabel
	\let\marginnote=\draftmarginnote
   \def\@eqnnum{(\theequation)\rlap{\kern\marginparsep\tt\@eqnlabel}%
\global\let\@eqnlabel\@vacuum}  }

\def \ba {\begin{eqnarray}}
\def \ea {\end{eqnarray}}

\newcommand{\nn}{\nonumber}

\newcommand{\figket}[2][0.3]{\Bigl|\,%
\raisebox{-0.9ex}{\includegraphics[scale=#1]{#2}}\,%
\Bigr\rangle}
\newcommand{\figbra}[2][0.3]{\Bigl\langle \,%
\raisebox{-0.9ex}{\includegraphics[scale=#1]{#2}}\,%
\Bigr|}
\newcommand{\configA}{\Bigl|\,%
\raisebox{-1.0ex}{\includegraphics[scale=0.3]{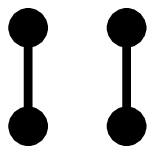}}\,%
\Bigr\rangle}
\newcommand{\configB}{\Bigl|\,%
\raisebox{-1.0ex}{\includegraphics[scale=0.3]{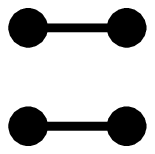}}\,%
\Bigr\rangle}
\newcommand{\configHuno}{\Bigl|\,%
\raisebox{-1.0ex}{\includegraphics[scale=0.3]{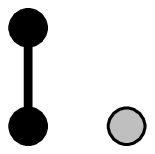}}\,%
\Bigr\rangle}
\newcommand{\configHtres}{\Bigl|\,%
\raisebox{-1.0ex}{\includegraphics[scale=0.3]{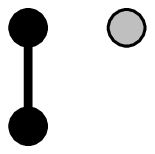}}\,%
\Bigr\rangle}
\newcommand{\configHcinco}{\Bigl|\,%
\raisebox{-1.0ex}{\includegraphics[scale=0.3]{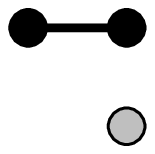}}\,%
\Bigr\rangle}
\newcommand{\configHsiete}{\Bigl|\,%
\raisebox{-1.0ex}{\includegraphics[scale=0.3]{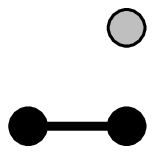}}\,%
\Bigr\rangle}
\newcommand{\braA}{\Bigl\langle \,%
\raisebox{-1.0ex}{\includegraphics[scale=0.3]{estadoA.eps}}\,%
\Bigr|}
\newcommand{\braB}{\Bigl\langle \,%
\raisebox{-1.0ex}{\includegraphics[scale=0.3]{estadoB.eps}}\,%
\Bigr|}
\newcommand{\braHdos}{\Bigl\langle \,%
\raisebox{-1.0ex}{\includegraphics[scale=0.3]{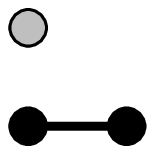}}\,%
\Bigr|}
\newcommand{\braHcuatro}{\Bigl\langle \,%
\raisebox{-1.0ex}{\includegraphics[scale=0.3]{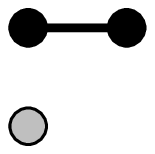}}\,%
\Bigr|}
\newcommand{\braHseis}{\Bigl\langle \,%
\raisebox{-1.0ex}{\includegraphics[scale=0.3]{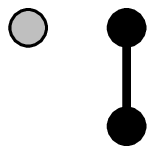}}\,%
\Bigr|}
\newcommand{\braHocho}{\Bigl\langle \,%
\raisebox{-1.0ex}{\includegraphics[scale=0.3]{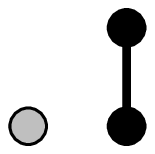}}\,%
\Bigr|}
\newcommand{\sumauno}{%
\raisebox{0.0ex}{\includegraphics[scale=0.15]{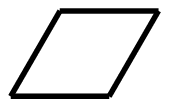}}
}
\newcommand{\sumados}{%
\raisebox{0.0ex}{\includegraphics[scale=0.15]{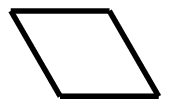}}
}
\newcommand{\sumatres}{%
\raisebox{0.0ex}{\includegraphics[scale=0.15,angle=60]{suma1.eps}}
}

\def\prl#1#2#3{Phys.\ Rev.\ Lett.\ {\bf #1}, #2 (#3)}

\def\prb#1#2#3{Phys.\ Rev.\ B {\bf #1}, #2 (#3)}

\def\advphys#1#2#3{Adv.\ in Phys.\ {\bf #1}, #2 (#3)}

\def\science#1#2#3{Science {\bf #1}, #2 (#3)}

\def\euro#1#2#3{Euro.\ Phys.\ Lett.\ {\bf #1}, #2 (#3)}

\begin{document}

\title{Statistical transmutation in doped quantum dimer models.}

\author{C.A. Lamas}
\affiliation{Laboratoire de Physique Th\'eorique,  IRSAMC,
 CNRS and Universit\'e de Toulouse, UPS, F-31062 Toulouse, France}

\author{A. Ralko}
\affiliation{Institut N\'eel, CNRS and Universit\'e Joseph Fourier, F-38042 Grenoble, France
}

\author{D.C. Cabra}
\affiliation{IFLP, Departamento de F\'isica, Universidad Nacional de La Plata, La Plata, Argentina.}

\author{D. Poilblanc}
\affiliation{Laboratoire de Physique Th\'eorique,  IRSAMC,
 CNRS and Universit\'e de Toulouse, UPS, F-31062 Toulouse, France}

\author{P. Pujol}
\affiliation{Laboratoire de Physique Th\'eorique,  IRSAMC,
 CNRS and Universit\'e de Toulouse, UPS, F-31062 Toulouse, France}

\begin{abstract}
We prove a ``statistical transmutation" symmetry of doped quantum dimer models
on the square, triangular and kagome lattices: the energy spectrum is invariant under a {\it simultaneous} change of statistics (i.e. bosonic into fermionic or vice-versa) of the holes
and of the signs of all the dimer resonance loops.
This exact transformation enables to define duality equivalence between doped
quantum dimer Hamiltonians,
and provides the analytic framework to analyze
dynamical statistical transmutations.
We investigate numerically the doping of the triangular quantum dimer model,
with special focus on the topological $\mathbb{Z}_2$ dimer liquid.
Doping leads to four
(instead of two for the square lattice)
inequivalent
families of Hamiltonians. Competition between phase separation, superfluidity, supersolidity and fermionic phases
 is investigated in the four families.
\end{abstract}
\maketitle

The discovery of exotic liquids such as topological $\mathbb{Z}_2$
spin liquids (SL) \cite{Wen,Moessner_prl_2001_Z2} is one of the challenges of modern condensed matter physics.
Anderson proposed that the parent (insulating) state of the high-temperature
superconductors is in fact a SL, the resonating valence bond (RVB) state and that
spin-charge separation  (and superconductivity) will occur under doping \cite{AndersonRVB}:
the original electron fractionalizes into two emergent particles, a holon carrying the charge quantum
and a spinon carrying the spin quantum. Although the original electron is a fermion, there has been
a long-standing debate regarding the actual statistics of holons and spinons
in such a ``deconfinement'' scenario. In this context, Rokhsar and Kivelson~\cite{RK} introduced the quantum dimer model (QDM)
as a simple effective model to describe magnetically disordered phases. The basis assumption
here is that the spinon spectrum is gapped \cite{Read+sachedv_1991} (strictly speaking, in the QDM it is
infinite) and the dimers between nearest neighbor (NN) sites mimic fluctuating SU(2) singlets of paired electrons.
The underlying microscopic exchange interaction leads to effective
attraction/repulsion between dimers and dimer-flips along closed loops (see below).
Doped QDM, where dimers (i.e. pairs of electrons) are removed from the system, leading to
itinerant holes, have also been studied \cite{RK,Kivelson_1989, RMP0, poilblanc_2008}.
There, holes and dimers are strongly coupled due to hard-core constraints.
Variants of doped QDM have been also constructed to physically describe polarized spinons induced by a magnetic field~\cite{RMP}.
 Naively, one expects holons to be of fermionic nature
 (while spinons should be bosonic) but, in spite of this naive expectation,
 the statistics of holons has been debated over the last 20 years.
Earlier work suggested that holon excitations were fermions~\cite{Read_1989}.
 It was then argued that, in fact, holon statistics is dictated by energetic considerations
 and under some conditions a holon could become a bosonic composite through binding
 of a flux quantum~\cite{Kivelson_1989}, the ``vison" in the QDM language.
 This was indeed observed recently by exact diagonalization techniques in the square
 lattice~\cite{poilblanc_2008} where, by varying the ratio of dimer kinetic energy vs holon kinetic energy,
one can transit from a regime where low-energy quasi-particles behave as fermions to a regime where they behave as bosons.

\begin{figure}[t!]
 \begin{center}
 \includegraphics[width=0.4\textwidth]{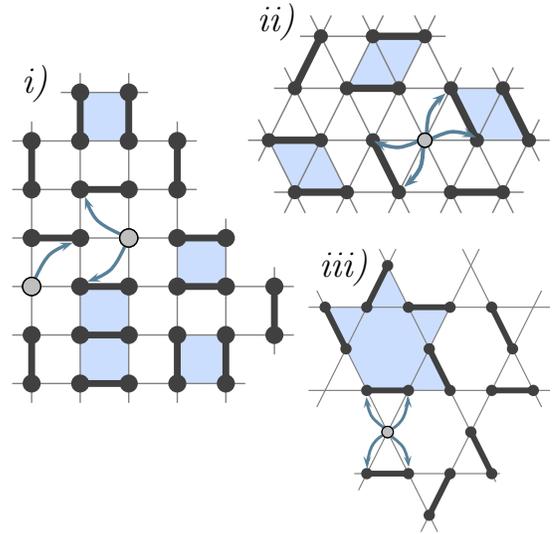}
\end{center}
\caption{\label{fig:fig_varias}
(color online) Dimer  coverings corresponding to
 different lattices. Light blue areas represent flippable plaquettes and  arrows represent the allowed single hole hopping process.
 i) Square lattice: The hole hopping is between sites in the same sublattice.
ii) Triangular lattice: A hole can hop in the different directions schematized by the arrows.
 iii) kagome lattice:
 The light blue area shows one of the possible flippable plaquettes of length 10.  }
\end{figure}

In this letter we prove an exact duality transformation that provides
a framework to analyze this dynamical statistical transmutation.
More precisely, we study the generic Hamiltonians of QDM's doped with holes
depicted in Fig. \ref{fig:fig_varias}, which contain three contributions, $H_J$ and $H_V$ respectively the flipping and potential energy terms for the dimers of amplitudes $J$ and $V$ and $H_t$ the hopping term for vacancies (named ``holes") of amplitude $t$.
%
The Hilbert space corresponds to all dimer coverings with a fixed hole density.
The potential term is proportional to the number of
flippable plaquettes (which is not conserved by dimer flips and hole motion).
Special attention is devoted to frustrated lattices,  the
triangular lattice with {\it edge-sharing} triangular units and  the kagome
lattice with {\it corner-sharing} triangular units, which we compare to the
results obtained for the square lattice~\cite{poilblanc_2008}.
The exact symmetry of these Hamiltonians states that their spectrum is
invariant under {\it simultaneous}
transformations of the statistics of the holes, bosons into fermions or vice-versa, ${\cal B}\leftrightarrow{\cal F}$,
and of the sign of (all) dimer kinetic amplitudes, $J\leftrightarrow -J$, $V$ and $t$ being unchanged.

\begin{figure}[h!]
 \begin{center}
  \includegraphics*[angle=-90,width=0.45\textwidth]{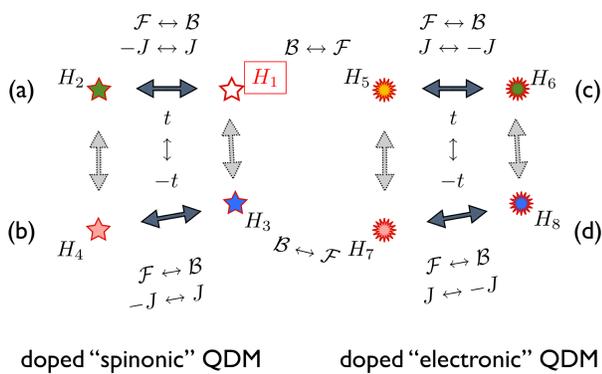}
\end{center}
\caption{
\label{fig:equiv_Hamiltonians}
(color online) Equivalence classes ("models") and relations between the eight Hamiltonians introduced in the text.
 Starting form the unfrustrated ($J>0$, $t>0$) bosonic $H_1$ Hamiltonian,
one can combine any change of statistics and/or change of signs of $J$ and/or sign of $t$ to obtain
$H_2,\cdots,H_8$. Equivalence relations are shown by arrows: full dark grey arrows
(dotted light grey arrows) are valid for all lattices (for vacancies hopping on the square lattice only).
}
\end{figure}

{\it Equivalence classes --} For fixed values of the magnitudes of $J$, $V$ and $t$, one can define eight different
 Hamiltonians by changing the signs of $J$ and/or $t$ and by
choosing bosonic or fermionic statistics for the mobile holes.
Using the exact symmetry mentioned above and proven below,
these eight different
 Hamiltonians can be grouped in pairs with identical spectra, hence
 defining four non-equivalent families.
 This is summarized in Fig. \ref{fig:equiv_Hamiltonians}.
More precisely, the established equivalence is between $H_1$ and $H_2$, both represented by the family label (a) and in the same way for
 the pairs of Hamiltonians $H_{2i-1}$ and $H_{2i}$ with $i=2,3,4$ corresponding respectively
 to families (b), (c) and (d).
 We argue that such an equivalence is a general feature of doped-two-dimensional QDMs, independently of the lattice geometry.
 Physically, classes (a) and (b) ((c) and (d)) describe systems with doped polarized spinons (doped holons), and are called ``spinonic'' QDMs (``electronic'' QDMs). Note that, for the square (or hexagonal) lattice,
 an additional $t \leftrightarrow -t$ symmetry reduces the number of non-equivalent families
 from 4 to 2.
 In Fig. \ref{fig:equiv_Hamiltonians},  the two families defined for the square lattice by
 $H_1\leftrightarrow H_2\leftrightarrow H_3\leftrightarrow H_4$
on one hand, and by $H_5\leftrightarrow H_6\leftrightarrow H_7\leftrightarrow H_8$
on the other hand, were dubbed as ``Perron-Frobenius" and ``non-Perron-Frobenius"
Hamiltonians and studied in \cite{poilblanc_2008} while, for the
 triangular lattice, only the unfrustrated $H_1$ Hamiltonian (representing family (a))
 was studied in \cite{RMP}.

As the transformation we shall establish below change bosons into fermions
(and vice versa), we can start by
 assuming,  without lost of generality, that the (bare) holons  in the doped QDM are bosons.
 We implement a two dimensional Jordan-Wigner transformation on these bosons to change their statistics to fermionic.
In  contrast to one dimensional systems,
this resulting transformed Hamiltonian is
highly non-local \cite{Fradkin_prl_1989,Wang_1991,Cabra_JW,Lamas_2006} requiring,
in general,
a mean field approximation to proceed further, at least from an analytical point of view.
Next, we show that the non-local terms can be absorbed by using
a different representation for the dimer operators, which keeps their bosonic character.
 This key feature provides then an elegant proof of the ``statistical
 transmutation symmetry" of these models.

{\it Proof of statistical transmutation symmetry --}
It is convenient to write the Hamiltonian in a second quantized form, by introducing creation operators $b^{\dagger}_{i,j}$ for a dimer siting between sites $i$ and $j$ and holes by operators
$a^{\dagger}_{i}$. In our conventions, dimers between sites $i$ and $j$ are created by spatially
symmetric operators $b^{\dagger}_{i,j}$, both operators ($b^{\dagger}_{i,j}$ and $a^{\dagger}_{i}$) are bosonic and mutually commuting.
It is instructive to notice that the dimer bosonic operators can be thought as bilinears of ``electrons'' operators:
$b^{\dagger}_{i,j}=\frac{1}{\sqrt{2}}(c^{\dagger}_{i,\uparrow}c^{\dagger}_{j,\downarrow}-c^{\dagger}_{i,\downarrow}c^{\dagger}_{j,\uparrow})$.
In terms of these operators we implement the hard-core constraint
 $a^{\dagger}_{i}a_{i}+\sum_{z}b^{\dagger}_{i,i+z}b_{i,i+z}=1,$
where the sum runs over NN of site $i$.
Let us call $\hat{\mathcal{P}}$ the projector on the subspace where the constraint
 holds. In the following we use systematically the projected Hamiltonian
$
\hat{\mathcal{P}} H \hat{\mathcal{P}}
$
($=H \hat{\mathcal{P}}$ since all terms of $H$ commute with $\hat{\mathcal{P}}$),
which prove to be very useful later.

Let us now apply 2-D Jordan-Wigner (JW) transformation on the holon operators \cite{Wang_1991}:
\ba
\label{eq:Jordan_Wigner}
a_{i}=e^{-i\phi_{i}}f_{i}
\ea
where
$
\phi_{i}=\sum_{j\neq i}f^{\dagger}_{j}f_{j} \arg(\tau_{j}-\tau_{i})
$
and $\tau_{j}=x_{j}+iy_{j}$ is the complex coordinate of the j-th hole.
Using that $\arg(\tau_{i}-\tau_{j})=\arg(\tau_{j}-\tau_{i})\pm\pi$
it follows immediately that two $f$-operators in different sites anticommute and the constraint
 and the phase $\phi_i$ can be written equally in terms of
$f_{i}$ or $a_{i}$ operators.

In order to understand the consequences of  transformation  (\ref{eq:Jordan_Wigner}) on the Hamiltonian,
the hopping of holons can be written, for an arbitrary lattice,
as a sum of three-site Hamiltonians
\ba
\nn
H_{t}=\sum h^{(t)}_{(ijk)}
~~\hbox{with}~~
\label{eq:hoping_gral}
h^{(t)}_{(ijk)}=t\;\hat{\mathcal{P}}\;b^{\dagger}_{i,j}b_{j,k}a^{\dagger}_{k}a_{i}\;\hat{\mathcal{P}}.
\ea
Making use of the transformation (\ref{eq:Jordan_Wigner}) we  obtain
\ba
\nn
h^{(t)}_{(ijk)}=t\;e^{i\arg(\tau_{k}-\tau_{i})}\hat{\mathcal{P}} e^{i\phi_{k}}e^{-i\phi_{i}}
\;b^{\dagger}_{i,j}b_{j,k}
f^{\dagger}_{k}f_{i}\;\hat{\mathcal{P}}.
\ea %
In the last equation we have changed boson operators $a_{i}$ by fermionic
ones $f_{i}$ at the cost of introducing non-local interactions.
In other words we have written a boson as a composite particle consisting of an
electron with an attached flux.
As we show just below, this non-local interaction can be absorbed in a redefinition of
the dimer operators without affecting their bosonic commutations relation.
%
%
In order to define new dimer operators including the phases $\phi$ in their definition, we must be able
to write $\phi$  in terms of operators $b_{i,j}$. This can be performed in the following way:
When applied into the subspace projected by $\hat{\mathcal{P}}$, we can change $\phi$ by $\tilde{\phi}$ in the exponentials, where
\ba
\nn
\tilde{\phi}_{i}=\sum_{r\neq i}\left[1-\sum_{z}b^{\dagger}_{r,r+z}b_{r,r+z}\right] \arg(\tau_{r}-\tau_{i}).
\ea
The remaining non-local exponential operators are re-written in terms of $\tilde{\phi}$
 which can be absorbed by defining $\tilde{b}_{i,j}= e^{i(\tilde{\phi}_{i}+\tilde{\phi}_{j})} \, b_{i,j}$.
As announced above, it is a simple matter to see that operators $\tilde{b}_{i,j}$ are also bosonic, as it should be to make sense
 as dimer operators.
Then, transformation (\ref{eq:Jordan_Wigner}) together with the definition of $\tilde{b}_{i,j}$
 allow us to change the statistics of holes from bosonic to fermionic. After this transformation
the hopping term is written in terms of operators $\tilde{b}_{i,j}$ and $f_{i}$ and the hopping
 amplitude changes to $\tilde{t}=t\,e^{i[\pi+\arg(\tau_{j}-\tau_{i})-\arg(\tau_{j}-\tau_{k})]}$.
Let us now investigate the effect of this transformation specifically for each lattice~:

{\it i) Square lattice. -- }
 The Hamitonian is defined by $H_{1}=H_{J}+H_{V}+H_{t}$, where:
\ba
\nonumber
H_{J}&=&-J\sum_{\Box}  \left\{\configA \braB + \hbox{H.C.}\right\}\\
\nonumber
H_{V}&=&V\sum_{\Box} \left\{\configA \braA + \configB \braB  \right\}\\
\nonumber
H_{t}&=&t\sum_{\Box}\left\{
\configHuno \braHdos + \configHtres \braHcuatro \right.\\ \nonumber
&+&\left. \configHcinco \braHseis + \configHsiete \braHocho + \hbox{H.c.}
\right\}.
\ea
Before going to the details of the Jordan-Wigner transformed Hamiltonian let us mention some well
 known gauge transformations
which can be performed. For zero doping,
 a simple gauge transformation on the dimers can be done to show the equivalence of the Hamiltonians with
$J$ and $-J$. In the case of non-zero doping, changing the sign of the holes wave function with a
 wave vector transformed as $\vec{k}\rightarrow \vec{k}+ (\pi, \pi)$ one obtains easily the equivalence between Hamiltonians with $t$ and $-t$.
We can now proceed to the computation of the Jordan-Wigner transformed Hamiltonian.
The transformed Hamiltonian becomes,  after some algebra \cite{SM},  $H_{2}=\tilde{H}_{J \to -J}+\tilde{H}_{V}+\tilde{H}_{t}$,
where the tilde means that dimers  and holes are created by
operators $\tilde{b}^{\dagger}$ and $f^{\dagger}$ respectively.
 In the dimer kinetic Hamiltonian the amplitude $J$ has been changed by $-J$.


{\it ii) Triangular lattice. -- }
Here the Hamiltonian $H_{1}$ is written using:
\ba
\nonumber
H^{(J)}_{\sumauno}&\!\!\!=&\!\!-J\sum_{\sumauno}  \left\{\figket{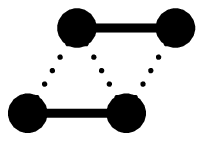} \figbra{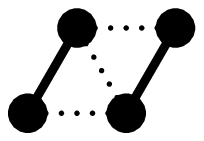}+ \hbox{H.c.}\right\}\\
\nonumber
H^{(V)}_{\sumauno}&\!\!\!=&\!\!V\sum_{\sumauno} \left\{\figket{flipTriangA.eps}\figbra{flipTriangA.eps}
 + \figket{flipTriangB.eps}\figbra{flipTriangB.eps}  \right\}\\\nn
%
%
H^{(t)}_{\triangle}&\!\!\!=&\!\!t\sum_{\vartriangle}\!\left\{
\figket{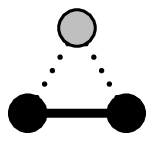}\figbra{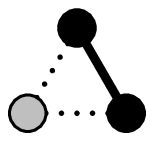}
\! +\! \figket{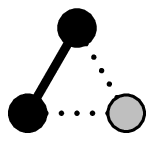}\figbra{h1.eps}
\!+\! \figket{h2.eps}\figbra{h3.eps}
 \!+\! \hbox{H.c.}\!  \right\}
\ea
and similar expressions for $H_{\sumados}$ and $H_{\sumatres}$ for the
other orientations of the rhombi \cite{Albuquerque}.
$H^{(t)}_{\triangle}$ corresponds to the holon hopping on the
up-triangles, the total hopping Hamiltonian being
 $H_{t}=H^{(t)}_{\triangle}+H^{(t)}_{\triangledown}$.  It is important to
note that, on such non-bipartite
lattices, the equivalence $t
\leftrightarrow -t$ does not hold anymore, as verified
numerically.
This implies that there are now four families of equivalent
Hamiltonians, instead of two for the square case.
On the triangular lattice, the dual Hamiltonian reads  $H_{2}=\tilde{H}^{(J)
\to (-J)}_{\sumauno}+\tilde{H}^{(V)}_{\sumauno}+\tilde{H}^{(t)}$, with the same
convention as before \cite{SM}.
A corollary of this result is that, as for the square
lattice, the Hamiltonians with $J$ and $-J$ are equivalent  in the
zero doping case.

 {\it iii) Kagome lattice. --}
 On this geometry, all dimer resonant loops (of length $\alpha=6, 8, 10, 12$)
involving a single hexagon become important and should be considered~\cite{PMS}.
Nearest-neighbor hole hopping as depicted in Fig. \ref{fig:fig_varias} can be
introduced as in \cite{Ralko_Poilblanc_kagome} (the reader can refer
to \cite{PMS,Ralko_Poilblanc_kagome} for more details).
The JW transformed Hamiltonian is obtained in the same way as
before and since it is interesting on its own, it will be presented elsewhere
\cite{large}. Here, we only summarize the main results.
By using the JW transformation followed by (more involved) gauge transformations on dimer and hole operators,~\cite{note:gauge_transformation} one can  write
 the dual Hamiltonian with all kinetic dimer amplitudes $J_{\alpha}$ changed
 into $-J_{\alpha}$.
 As before, dimers and holons are created by  $\tilde{b}^{\dagger}$ and $f^{\dagger}$ respectively.

{\it Statistical transmutation and choice of Hamiltonian --} The above statistical transmutation symmetry is of great importance in analytic and numerical investigations of doped QDMs, establishing their phase diagrams.
In \cite{poilblanc_2008}, it was shown that,
in the square lattice, by varying the ratio $J/t$ or the doping, the system undergoes a series of phase transitions. One of those phase transitions corresponds to a dynamical transmutation,
between a  phase where elementary low-energy
quasi-particles are fermionic holes, to a phase where the low-energy quasi-particles
acquire a bosonic nature.
\begin{figure}
 \begin{center}
\includegraphics*[angle=0,width=0.5\textwidth]{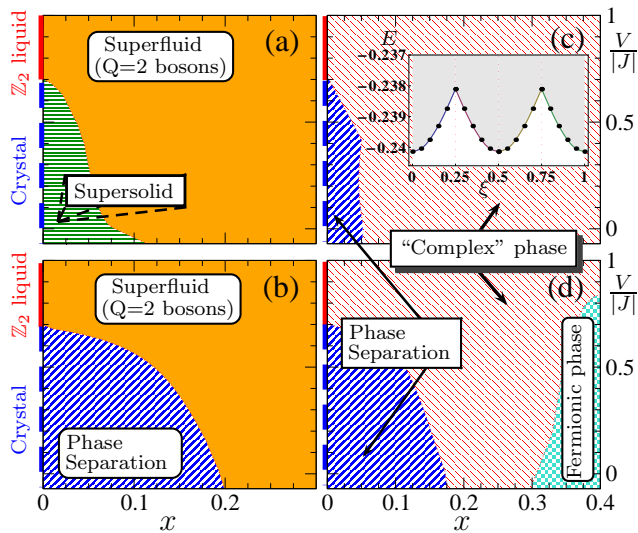}
\end{center}
\caption{
\label{fig:phasediag_triangle}
(color online) Qualitative phase diagrams of the four families of doped QDM's (a-d)
as labeled in Fig.~\ref{fig:equiv_Hamiltonians} on the triangular lattice
versus doping ($x$) and $V /|J|$ for fixed $t/J=0.5$. Inset: Energy vs normalized
magnetic flux ($\xi$)
corresponding to family (c) for fixed  $V /|J|=0.5$,  $t/J=0.5$ and $x=0.25$. The $E$ vs $\xi$ plots
corresponding to families (a), (b) and (d) are qualitatively similar.
}
\end{figure}
It was in fact predicted that holon and vison could pair up, leading to such statistical transmutation~\cite{Kivelson_1989}.
Then, if the microscopic Hamiltonain is chosen such that holons are fermions, the
fermionic phase corresponds to a weak-coupling regime and the bosonic phase to a strong-coupling one. This picture gets interchanged if holons in the microscopic
 Hamiltonian are chosen to be bosons.
 Hence, using this duality equivalence one can always choose the most relevant
 microscopic QDM Hamiltonian to be in a weak-coupling regime (depending on the point in the phase diagram).

{\it Phase diagrams --} We now complement the exact results with a numerical study of the phase diagram of the four non-equivalent families of
 Hamiltonians in the triangular lattice where a topological ($\mathbb{Z}_{2}$) liquid
 can be stabilized at zero doping \cite{Moessner_prl_2001_Z2,Ralko_2005}.
The statistics of the dressed excitations is studied using the method developed in \cite{poilblanc_2008} by investigating the node content of the wave functions.
Family (a) in Fig. \ref{fig:phasediag_triangle} is the only unfrustrated case and was studied with Green Function Monte Carlo methods in \cite{RMP}. In this model bare holons are bosonic and remain representative of the physical excitations in the entire region of the phase diagram, as happen also in the Perron-Frobenius square lattice version, studied  in \cite{poilblanc_2008}.
The unfrustrated model shows a superfluid phase in a wide range of the phase diagram.

The situation is even more interesting when $J$ is changed into $-J$, or equivalently, bosons
are changed into fermions (families (c) and (d) in Fig. \ref{fig:phasediag_triangle}). Most of the phase diagram is occupied by what we dubbed a ``complex phase" as the statistics of dressed excitations does not correspond solely to bosons or fermions. Similarly
to the square lattice, we find a phase separated region at low doping.
Family (d) also shows, in addition, an interesting fermion reconstruction at large doping,
which we call ``fermionic phase".

In order to provide a better understanding of the conducting phases of the
triangular lattice, we insert in the torus an Aharonov-Bohm flux of strength
$\phi = \xi \phi_0$ with $0 \le \xi \le 1$
(achieved by adding a phase shift in e.g. the x-direction)
 and where $\phi_0 = hc / e$ the
elementary magnetic flux. As done in Ref.~\cite{RMP0} on the square lattice,
one can reduce the finite-size effects by considering arbitrary boundary conditions in the
$y$-direction.
A superfluid is characterized by well-defined minima in the
ground state energy
separated by a finite barrier in the thermodynamic limit. A
contrario, a typical signature of (weakly interacting) fermions, a flat energy
profile is expected even on such a small cluster \cite{Poilblanc1991}. Here, we
report that the ground state energy
has well-defined minima quantized at half a flux quantum for all family of models at $x\sim 0.25$,
compatible with a bare charge $Q=2 e$ superconductor~\cite{2e, RMP0}(see inset of Fig.~3(c)).
Further detailed discussions on the charge of the ``low energy" particles in the superconducting phase will be provided elsewhere \cite{large}.

{\it Discussion and perspectives --}
 In this paper we have shown that the nature of the excitations in doped QDM is a much more subtle question than
what one would naively think. We have rigorously established equivalence
classes between QDM Hamiltonians (see Fig. \ref{fig:equiv_Hamiltonians}).
We claim that this duality relation is
a generic feature of QDM independently of the details of the lattice.

This proven duality relation provides a powerful tool to identify the nature of the
dressed hole excitations. Indeed, imagine one considers a specific system, with given microscopic parameters, 
corresponding to one of the four (or two) families of Hamiltonians.
{\it A priori} two (or four) representations of this system can be chosen freely 
using the equivalence relations of Fig.~\ref{fig:equiv_Hamiltonians}.
As we have shown, there is in fact a specific representation which corresponds to 
a weakly coupled regime, {\it i.e.} where the chosen statistics of the bare holes {\it is} the one of the true dressed
excitations. This is the representation to be chosen to study {\it e.g.} the
effect of further perturbations of the system. 


Lastly, we have shown numerically that models on the triangular lattice
corresponding physically to spinon or holon doping have very different phase diagrams with either a superfluid or a ``complex" phase (in which dressed excitations can not be understood solely in terms of fermionic or bosonic degrees of freedom), respectively.
In the latter case, we provide evidence of
statistical transmutation, analogously to the case of the square lattice.

{\it Acknowledges ---}
AR, DP and PP  acknowledge support by the ``Agence Nationale de la Recherche" under grant No.~ANR~2010~BLANC~0406-0.
CAL and DCC are partially supported by CONICET (PIP 1691) and ANPCyT (PICT 1426).


\begin{thebibliography}{breitestes Label}


\bibitem{Wen} X.-G.Wen, \advphys{44}{405}{1995} and references therein.

\bibitem{Moessner_prl_2001_Z2} R.\ Moessner and S. L.\ Sondhi, \prl{86}{1881}{2001}

\bibitem{AndersonRVB} P. W. Anderson, \science{235}{1196}{1987}; P.W.~Anderson, Mater. Res. Bull. {\bf 8}, 153 (1973);
P.~Fazekas and P.W.~Anderson, Philos. Mag. {\bf 30}, 23 (1974).

\bibitem{RK} D. S.\ Rokhsar and S. A.\ Kivelson, \prl{61}{2376}{1988}.

\bibitem{Read+sachedv_1991} N. Read and S. Sachdev, \prl{66}{1773}{1991}.

\bibitem{RMP0}
A. Ralko, F. Mila and D.\ Poilblanc, \prl{99}{127202}{2007};
see also D.~Poilblanc et al., Phys.Rev. B {\bf 74}, 014437(2006).

\bibitem{poilblanc_2008} D.\ Poilblanc, \prl{100}{157206}{2008}.

\bibitem{Kivelson_1989} S.\ Kivelson, \prb{39}{259}{1989}.

\bibitem{RMP} A. Ralko, F. Becca and D. Poilblanc, Phys. Rev. Lett. {\bf 101}, 117204 (2008);

\bibitem{Read_1989} N.\ Read and B.\ Chakraborty, \prb{40}{7133}{1989}.

\bibitem{Fradkin_prl_1989} Eduardo Fradkin, \prl{63}{322}{1989}.

\bibitem{Wang_1991} Y.R.\ Wang, \prb{43}{3786}{1991}.

\bibitem{SM} See Supplemental Material at [URL to be inserted by publisher] for the technical details.

\bibitem{Cabra_JW} D.C. Cabra, G.L. Rossini. \prb{69}{184425}{2004}.

\bibitem{Lamas_2006} C.A. Lamas, D.C. Cabra, M.D. Grynberg, G.L. Rossini. \prb{74}{224435}{2006}.

\bibitem{Albuquerque} A.F.\  Albuquerque, H.G.\ Katzgraber, M.\ Troyer, G.\ Blatter, \prb{78}{014503}{2008}
\bibitem{PMS} D. Poilblanc, M. Mambrini and D. Schwandt, \prb{81}{180402(R)}{2010}.

\bibitem{Ralko_Poilblanc_kagome} D.\ Poilblanc and A.\ Ralko, \prb{82}{174424}{2010}.

\bibitem{large} C.A.\ Lamas, A.\ Ralko, D.C.\ Cabra, D.\ Poilblanc, M.\ Oshikawa and P.\ Pujol, (unpublished)

\bibitem{note:gauge_transformation}
The gauge transformation allows to bring the hopping term back to its original
form after changes in the hopping amplitudes generated by the statistical transmutation.
For convenience, details will be published in \cite{large}.

\bibitem{Ralko_2005} A.\ Ralko, M.\ Ferrero, F.\ Becca, D.\ Ivanov and F.\ Mila, \prb{71}{224109}{2005}.

\bibitem{2e} S. A. Kivelson, D. S. Rokhsar and J. P. Sethna, \euro{6}{353}{1988}.

\bibitem{Poilblanc1991} D. Poilblanc, \prb{44}{9562}{1991}.




\end{thebibliography}
\end{document}